\documentclass[10pt,conference]{IEEEtran}
\IEEEoverridecommandlockouts
\usepackage{amsmath}
\usepackage{mathrsfs,amsthm,amsmath,amssymb}
\usepackage{textcomp}
\usepackage{graphicx,multirow}
\usepackage{balance}
\usepackage{cite}
\usepackage{gensymb}
\usepackage[center]{caption}

\usepackage{xcolor}

\DeclareMathOperator{\st}{s.t.}

\begin{document}

\title{\textcolor{black}{Energy-efficient Functional Split in  Non-terrestrial  Open Radio Access Networks} }
  \author 
{\IEEEauthorblockN{S. M. Mahdi Shahabi}
\IEEEauthorblockA{Department of Engineering\\
King's College London\\
London, UK\\
Email: mahdi.shahabi@kcl.ac.uk}
\and
\IEEEauthorblockN{Xiaonan Deng}
\IEEEauthorblockA{Department of Engineering\\
King's College London\\
London, UK\\
Email: xiaonan.deng@kcl.ac.uk}
\and
\IEEEauthorblockN{Ahmad Qidan}
\IEEEauthorblockA{Department of Engineering\\
King's College London\\
London, UK\\
Email: ahmad.qidan@kcl.ac.uk}
\and
\IEEEauthorblockN{Taisir Elgorashi}
\IEEEauthorblockA{Department of Engineering\\
King's College London\\
London, UK\\
Email: taisir.elgorashi@kcl.ac.uk}
\and
\IEEEauthorblockN{Jaafar Elmirghani}
\IEEEauthorblockA{Department of Engineering\\
King's College London\\
London, UK\\
Email: jaafar.elmirghani@kcl.ac.uk}
}
 \markboth{}{}
 \IEEEpubid{}
\maketitle

\begin{abstract}
This paper investigates the integration of Open Radio Access Network (O-RAN) within non-terrestrial networks (NTN),  and optimizing the dynamic functional split between Centralized Units (CU) and Distributed Units (DU) for enhanced energy efficiency in the network. We introduce a novel framework utilizing a Deep Q-Network (DQN)-based reinforcement learning approach to dynamically find the optimal RAN functional split option and the best NTN-based RAN network out of the available NTN-platforms according to real-time conditions, traffic demands, and limited energy resources in NTN platforms. This approach supports capability of adapting to various NTN-based RANs across different platforms such as LEO satellites and high-altitude platform stations (HAPS), enabling adaptive network reconfiguration to ensure optimal service quality and energy utilization. Simulation results validate the effectiveness of our method, offering significant improvements in energy efficiency and sustainability under diverse NTN scenarios.
\end{abstract}


\section{Introduction}\label{Introduction}

The deployment of the 5G technologies has initiated a transformative era in wireless networks, characterized by
unprecedented levels of connectivity, throughput, and low-latency communications. Non-Terrestrial Networks (NTN), which integrate satellite communication mechanisms into the 5G framework, play a pivotal role in extending these advanced network capabilities beyond the confines of terrestrial infrastructure. Recognized for their ability to deliver expansive global coverage, NTNs are essential for achieving the universal service goals of 5G\cite{azari2022evolution,3gpp2019study}. The 3rd Generation Partnership Project (3GPP) has embraced the integration of NTNs into the 5G New Radio (NR) standards, reflecting ongoing initiatives to harmonize satellite and ground-based network operations.

Parallel to the development of NTNs, the concept of Open Radio Access Networks (O-RAN) is redefining traditional network architectures. O-RAN promotes a more open, flexible, and scalable network infrastructure, which leverages disaggregation, virtualization, and programmability. This approach fundamentally alters how network entities like Radio Units (RU), Distributed Units (DU), and Centralized Units (CU) are deployed and managed, shifting away from the rigid, hardware-centric models of the past. The architectural overhaul provided by O-RAN facilitates a rapid, software-based reconfiguration of network operations, enhancing the network's ability to respond to dynamic conditions and varying demands. While the benefits of integrating O-RAN within NTNs are manifold, this fusion introduces complex challenges, particularly in the management of energy resources. In terrestrial settings, energy concerns, while important, are mitigated by relatively abundant power availability and robust infrastructure. However, in NTNs, especially those involving satellite components, power efficiency becomes a critical constraint due to limited energy resources available on satellites. This necessitates innovative approaches to manage and optimize energy consumption effectively.
The CU-DU interface utilizes the Space-to-Ground link based on the feeder link. While a static setup allows for the determination of an optimal functional split during the design phase, the dynamic nature of the NTN's architecture necessitates adaptive adjustments due to frequent structural changes \cite{rodrigues2023hybrid,zhu2022delay}. To meet this challenge, the NTN architecture employs near-real-time Radio Intelligent Controllers (RICs) based on O-RAN principles. These controllers facilitate proactive optimization of the functional split by gathering network status data and computing the most efficient configuration. This optimization aims to minimize energy consumption within the satellite's payload, considering factors such as user traffic, payload capabilities, available power, and the performance of the CU-DU physical feeder link \cite{campana2023ran}.
There are some existing works in the literature focusing on RAN functional split in TN and NTN architectures. In \cite{small2016small},
 the possible RAN split options for
network functions have been discussed according to the 3GPP guideline. The authors in \cite{larsen2018survey}
study the requirements stemming from each split option suggested by the 3GPP.
In \cite{murti2020optimal,amiri2023energy}
the optimization for virtual network function splitting and traffic routing has been done in terrestrial networks using machine-learning based methods. The authors in \cite{rihan2023ran} investigate the potential  split options for integrated TN and NTN architectures, whereas a comparative analysis for possible split options has not been provided.
\begin{table*}[htbp]
\vspace{10pt}
\caption{Potential functional split options between  DU and CU in O-RAN  for the RU traffic load of $\lambda_{RU}$ Mbps based on \cite{murti2020optimal}}
    \centering
    \begin{tabular}{c|c|c|c|c} 
Split Option $(o)$ & Functions in DU & Functionss in CU & Peak Traffic (Mbps) & Latency (ms) \\
\hline 0 & $f 1 {\rightarrow} f 2 {\rightarrow} f 3$  (PHY-RLC-MAC-PDCP) & None & $\lambda_{RU}$ & 30  \\
1 & $f 1 {\rightarrow} f 2$ (PHY-RLC-MAC)& $f 3$ (PDCP) & $\lambda_{RU}$ & 30 \\
2 & $f 1$ (PHY)& $f 2 {\rightarrow} f 3$ (RLC-MAC-PDCP) & $1.02 \lambda_{RU}{+}1.5$ & 2 \\
3 & None & $f 1 {\rightarrow} f 2 {\rightarrow} f 3$ (PHY-RLC-MAC-PDCP) & 2500 & 0.25
\end{tabular}
    \label{tab of spl}
\end{table*}

\begin{table*}[htbp]
\caption{The computational loads of the NTN node and the gateway for different split options}
    \centering
    \begin{tabular}{c|c|c} 
Split option (o) & $COMP_{i}^{o}$ & $COMP_{GAT}^{o}$  \\
\hline 0 & $COMP_{PHY}{+}COMP_{RLC}{+}COMP_{MAC}{+}COMP_{PDCP}$ & $0$ \\
1 & $COMP_{PHY}{+}COMP_{RLC}{+}COMP_{MAC}$ & $COMP_{PDCP}$ \\ 
2 & $COMP_{PHY}$ & $COMP_{RLC}{+}COMP_{MAC}{+}COMP_{PDCP}$ \\
3 & $0$ & $COMP_{PHY}{+}COMP_{RLC}{+}COMP_{MAC}{+}COMP_{PDCP}$ 
\end{tabular}
    
    \label{COMP_SAT,GAT}
\end{table*}

Taking the above into consideration, in this paper, the NTN architecture, built upon O-RAN principles is investigated, where after utilizing near-real-time RICs, a \textcolor{black}{capability of flexibility to different NTN-based RANs}   is further supported  to facilitate system-aware and proactive optimization of the functional split for different types of connectivity, i.e., LEO satellite-based RAN and high-altitude platform station (HAPS)-based RAN. The UE is able to connect to either of these platforms based on its requirement in terms of the maximum latency and demanding traffic requirements. This paper specifically addresses the challenge of optimizing the functional split in O-RAN configurations within NTNs to maximize energy efficiency. We explore a novel framework where the CU is located at the gateway, and the DU is onboard the NTN node, i. e., LEO satellite or HAPS. This setup forms the basis for our investigation into how network functions can be optimally divided between the CU and the DU under constraints related to power consumption, latency, traffic demands and computational loads. To do so, we propose a deep Q-network (DQN)-based reinforcement learning being able to follow dynamic environments undergoing different traffic and latency situations, and find a power-efficient solution for optimal RAN functional split option and an appropriate platform for NTN-based RAN. 

The remainder of the paper is structured as follows: Section II reviews  the system model being investigated throughout this paper in details. In section III, the problem formulation for optimizing the functional split and UE-to-RAN access is presented. In Section IV, we propose a DQN-based reinforcement learning method for solving the outgoing optimization coming from functional split options in different connectivity. In Section V, numerical analysis and performance simulation are presented for the proposed method. Finally, the conclusion is provided.
\begin{figure}[t]
    \centering
    \includegraphics[width=\columnwidth]{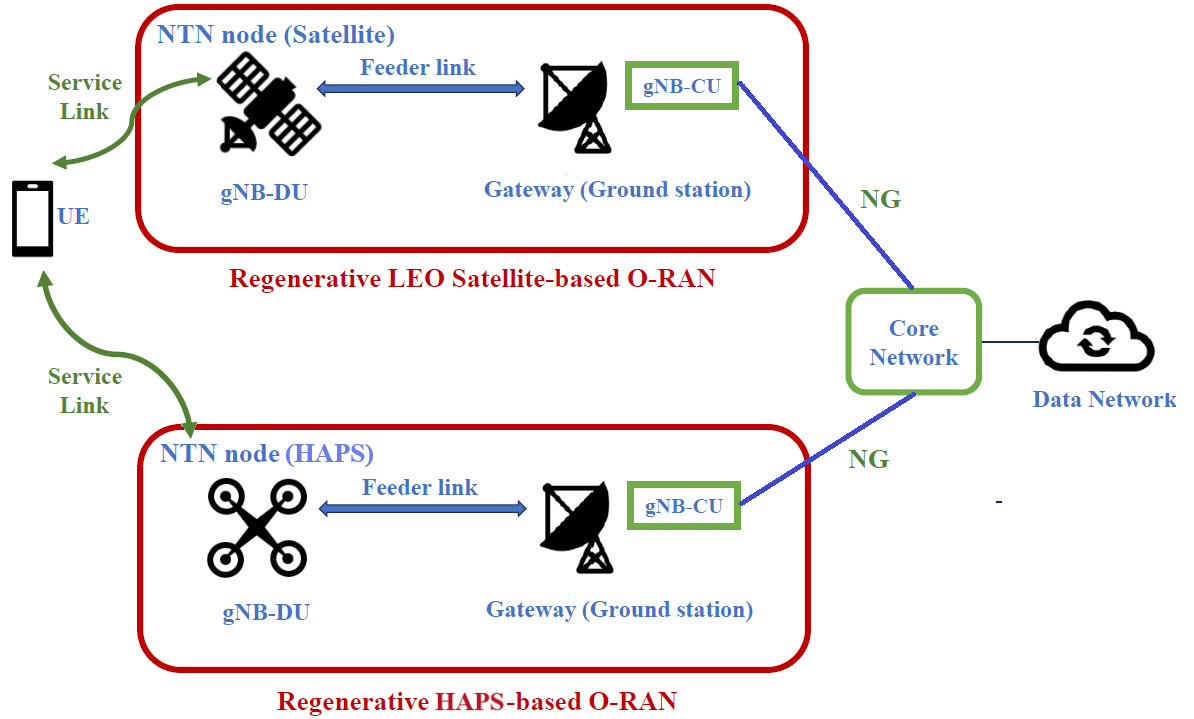}
    \caption{The overall NTN architecture}
    \label{Overall Arch}
\end{figure}
\section{System Model}
\begin{figure}[b]\label{splitoptions}
    \centering
    \includegraphics[width=0.6\columnwidth]{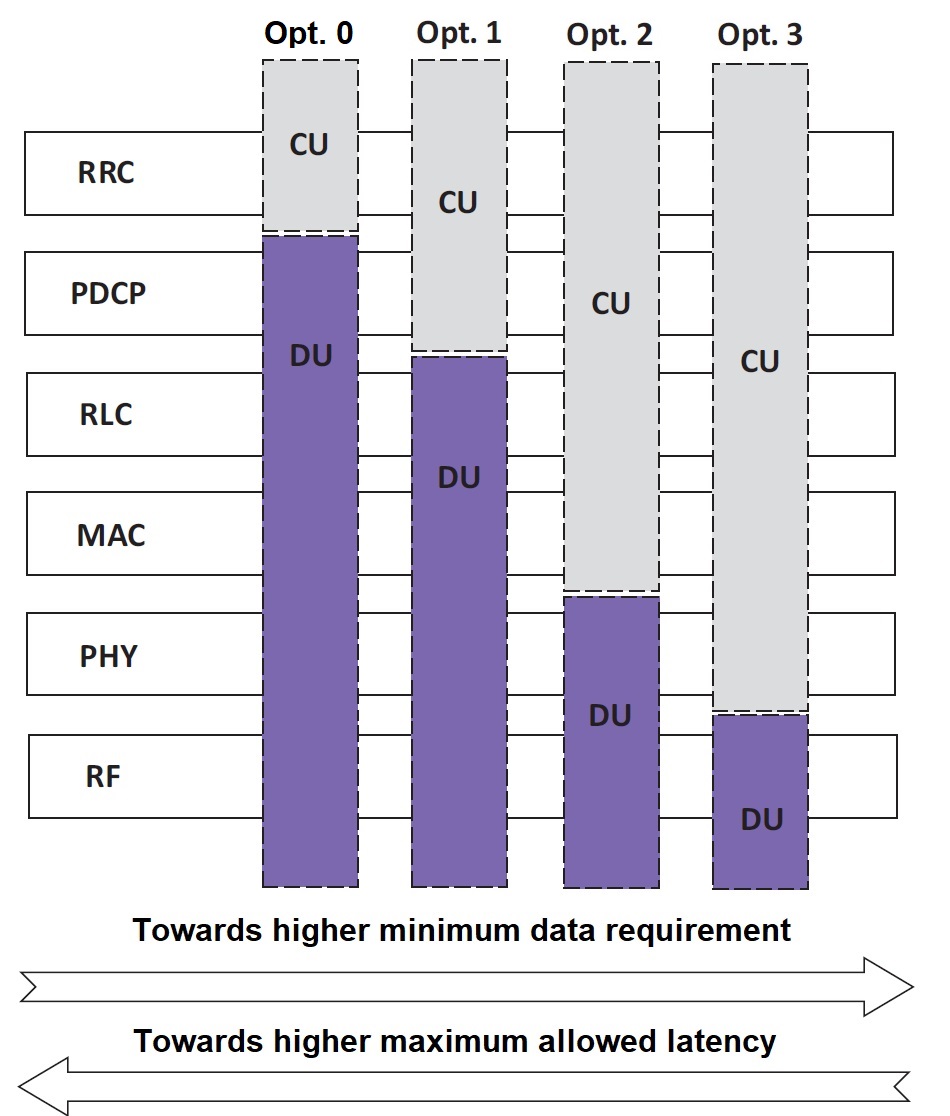}
    \caption{The trade-off between the data rate and the latency for RAN functional split options introduced by \cite{murti2020optimal}}
    \label{Overall Arch}
\end{figure}
The system model investigated in this paper consists of  NTN-based platform in which a UE is able to connects to a NTN node, i. e., LEO satellite or HAPS, through the service link. The NTN node on the other hand communicates with the NTN gateway which is connected to the 5G core network through the feeder link. The recent interfaces defined by the 3GPP enable the NTN nodes (LEO satellites and HAPSs) to operate as gNBs (NR logical nodes) partially or completely. Hence, the NTN nodes would be able to connect to UEs directly\cite{khan2023survey}. In this system model, based on the 3GPP TR 38.821\cite{3gpp2023solution}, we consider a  regenerative NTN-based O-RAN architecture  \textcolor{black}{being capable of adapting to different NTN-based RANs}  based on the traffic load and latency requirements of the UE, in which the gNB-CU is placed on on the ground station at the gateway while the gNB-DU and gNB-RU are located on the NTN nodes. Therefore, some processing tasks which are normally done by the gNB at the ground station or gateway can be performed by NTN node. This can results in network performance improvement in terms of meeting higher traffic demands\cite{bonafini2022analytical,wang2022seamless,cui2022space}. The overall network architecture is shown in Fig. 1.
The operation of the gNB is organized into a series of functions, with their distribution determined by functional splits. Adhering to the 3GPP framework, we utilize the four-option approach detailed in \cite{murti2020optimal} for  O-RAN. These options are summarized in Table I. We assume that in each time step $n$, a functional split option $o\in\mathcal{O}$ is selected from the possible split options in Table. I for each RAN network, where $\mathcal{O}=\{0,1,2,3\}$ is the set of possible RAN functional split options.
In this approach traffic flows are aggregated at RUs, DUs and CUs. In this case, a sequence of network functions $\{f0 \rightarrow f1 \rightarrow f2 \rightarrow f3\}$ is investigated, in which $f0$ should be placed in the RU driven by delay stringent requirements, while the other functions might be placed in either the CU or the DU as Virtual Machines (VMs). In this sequence, $f1$ denotes the PHY function, while $f2$  represents RLC and MAC functions, and and $f3$ stands for PDCP function. 
Increasing centralization results in enhanced resource management efficiency, thus lowering DU complexity and related expenses. However, this shift imposes stricter demands on data rates and latency, as depicted in Fig. 2. In other words, centralizing functions accelerates  the data load which should  be transferred to CUs. For instance, the data load increases from $\lambda_{RU}$ Mbps (payload) for $o=1$  to 2.5 Gbps for $o=3$. 
In the O-RAN architecture, the RU is connected to the DU  so that the DU serves the RU via Fronthaul (FH) link.
The DU which is usually located at the edge site is linked to the CU site via midhaul (MH) links to meet the delay required for the RAN split options. The  CU on the other hand is enabled with a direct link to the core network.

\section{Problem Formulation}
Prior to RAN functional split optimization, we should be able to calculate the key factors affecting the optimal split option, i. e., power consumption and latency. Therefore, in the following, we aim to obtain such parameters. 
In order to distinguish among potential split options, a binary  variable $y_{i}^{o} \in\{0,1\}$ is introduced to define deploying split $o \in \mathcal{O}$ for the NTN-based O-RAN network formed by the NTN node $i\in\mathbb{B}$ in each time slot, in which $\mathbb{B}=\{SAT,HAP\}$ is the set of NTN nodes, whose entries stand for the LEO satellite and the HAPS, respectively. In each time step, it is assumed the UE is connected to only one RAN network, and each RAN network selects one split option out of the possible options. In other words, we have

\begin{equation}\label{binarycond1}
    \sum_{o \in \mathcal{O}} y_{i}^{o}=1, \quad \forall i \in \mathbb{B},
\end{equation}
\begin{equation}\label{binarycond2}
    \sum_{i \in \mathcal{B}} y_{i}^{o}=1, \quad \forall o \in \mathcal{O}.
\end{equation}
\subsection{Power Consumption Calculation}
The total power consumption of the RAN functions for the split option $o\in\mathcal{O}$ can be obtained as the sum of the processing power $P_P^{o}$ and transmission power $P_T^{o}$ as follows
\begin{equation}\label{Total Power}
    P_{Total}^{o}=P_{P}^{o}+P_{T}^{o}.
\end{equation}
In (\ref{Total Power}), $P_{P}^{o}$ is the sum of the NTN node processing power and the gateway processing power for the split option $o$ as
\begin{equation}
    P_{P}^{o}=\left(\sum_{i \in \mathcal{B}}P_{P,i}^{o}y_{i}^{o}\right)+P_{P,GAT}^{o}.
\end{equation}
The values of $P_{P,i}^{o}, \ i\in\mathbb{B}$ and $P_{P,GAT}^{o}$ denote the processing power of the NTN node $i$ and the processing power of the gateway for the split option $o$ which  are calculated as follows
\begin{equation}\label{P_P_SAT}
    P_{P,i}^{o}=P_{I,i}+EPO_{i}.COMP_{i}^{o},
\end{equation}
\begin{equation}\label{P_P_GAT}
    P_{P,GAT}^{o}=P_{I,GAT}+EPO_{GAT}.COMP_{GAT}^{o},
\end{equation}
in which $P_{I,i}, \ i\in\mathbb{B}$ is the idle power of  processors/servers at the processing node $i$ and $EPO_{i}$ denotes the energy per operation for that node. Also, $P_{I,GAT}$ and $EPO_{GAT}$ represent the same parameters at the gateway processing node.    $COMP_{i}^{o}$ and $COMP_{GAT}^{o}$ in (\ref{P_P_SAT}) and (\ref{P_P_GAT}) are the  computational loads for the NTN node $i$ and the gateway, respectively, which  are given in the TABLE \ref{COMP_SAT,GAT}.

On the other hand, the transmission power $P_T^{o}$ in (\ref{Total Power}) for the split option $o$ is calculated by
\begin{equation}
    P_{T}^{o}=\sum_{i \in \mathcal{B}}P_{T,i}^{o}y_{i}^{o},
\end{equation}
in which
\begin{equation}\label{Transmission Power i}
    P_{T,i}^{o}=\frac{p_{i,GAT}}{C_{i,GAT}}TRA^{o}, \ i \in \mathcal{B}.
\end{equation}
In (\ref{Transmission Power i}), $C_{i,GAT}$ is the maximum link capacity between the NTN node $i$ and the gateway, $p_{i,GAT}$ denotes the power consumed by the NTN-node-i/gateway to transmit towards each other at the maximum link capacity and $TRA^{o}$ is the traffic demand of the RAN split option $o$ on the feeder link (the 4th column of TABLE. I) which is affected by the RU traffic $\lambda_{RU}$.
\subsection{Latency Calculation}
The Latency caused by RAN functional split mainly stems from the propagation of electromagnetic waves which is calculated by
\begin{equation}
    L_{Total}^{o}=\sum_{i \in \mathcal{B}}y_{i}^{o}\left(\frac{d_{i,GAT}}{c}\right),
\end{equation}
in which $d_{i,GAT}, \ \forall i \in \mathbb{B}$ denotes the distance between the NTN node $i$ and the gateway while $c$ represents the speed of the light.
\subsection{RAN functional Split Optimization Problem}
The main goal for the optimization of the RAN functional split option is to minimize the total power consumption while meeting the latency and traffic requirements. Hence, the resultant optimization problem would be as follows

\begin{equation}\label{main opt1}
\begin{array}{cl}
\min\limits_{\substack{y_{i}^{o} \\ i\in\mathbb{B} \\ o\in\mathcal{O}}}& P_{Total}^{o}\\ 
&    \\
\st 
&\displaystyle\sum_{o \in \mathcal{O}} y_{i}^{o}=1, \quad \forall i \in \mathbb{B},\\
&     \\
&\displaystyle\sum_{i \in \mathcal{B}} y_{i}^{o}=1, \quad \forall o \in \mathcal{O}, \\
&   \\
& L_{Total}^{o}\leq L^{o}, \ \forall i \in \mathbb{B}, \ \forall o \in \mathcal{O},\\
&    \\
& TRA^{o}\leq y_{i}^{o}C_{i,GAT}, \ \forall i \in \mathbb{B}, \ \forall o \in \mathcal{O},\\
&    \\
& y_{i}^{o}{COMP_{i}^{o}}\leq {COMP_{max,i}}, \ \forall i \in \mathbb{B}, \ \forall o \in \mathcal{O}, \\
&    \\
& {COMP_{GAT}^{o}}\leq {COMP_{max,GAT}}, \ \forall o \in \mathcal{O},\\

\end{array}
\end{equation}
in which $L^{o}$ stands for the latency requirement for the split option $o$ (in Table. \ref{tab of spl}). 
Also, $COMP_{max,i}$ and $COMP_{max,GAT}$ are the maximum computational capacity of the server at the NTN node $i$ and the maximum computational capacity of the server at the gateway in GOPS, respectively.
While the first and second constraints refer to  conditions in (\ref{binarycond1}) and (\ref{binarycond2}), the third constraint ensures that
the latency between the NTN nodes and the gateway does not exceed the maximum tolerable delay for the selected split option, and the fourth constraint ensures that the traffic resulting from the functional split  does not exceed the service link capacity. Also, the fifth and sixth constraints ensure that the computational loads would be below the computational capacity of the NTN nodes and the gateway. 
\section{DQN-based method}
In order to effectively respond to dynamic environments and facilitate efficient decision-making in addressing optimization problem (\ref{main opt1}), we recommend employing a deep reinforcement learning solution based on a DQN being able to characterize the dinamicity of the network against various traffic and latency requirements in a real-time manner. DQN as a reinforcement learning algorithm is employed for decision-making in dynamic environments, especially discrete environments, where there is a finite number of states and actions. It utilizes a Q-neural network (QNN) to approximate the optimal action-value function, enabling agents to learn optimal strategies through interactions with the environment through a sequence of state, action and reward. Agents select actions based on their current states, aiming to maximize cumulative rewards over time. Finally, rewards serve as feedback to the agent, guiding it towards optimal behavior and enabling adaptation to changing conditions. Therefore, prior to using DQN, it is crucial to effectively define the actions and reward being able to provide a correct system trajectory as well as the neural network required for the training process.
\subsection{State Set}
In each time step $n$, the state set is defined as follows
\begin{align}
    &s(n){=}\Big\{o(n),i(n),TRA^{o(n)},L^{o(n)},\lambda_{RU}(n),P_{Total}^{o(n)},L_{Total}^{o(n)},\nonumber\\
    &C_{i(n),GAT},
    COMP_{i(n)}^{o(n)},
    COMP_{GAT}^{o(n)},
    COMP_{max,i(n)}\Big\},
\end{align}
where $\lambda_{RU}(n)$, $o(n)$ and $i(n)$  denote the RU traffic load, the selected split option and the selected NTN platform in the time step $n$, respectively.
\subsection{Action Set}
Considering initial assumptions are made for the split option and the UE-to-RAN platform, i. e., LEO satellite or HAPS, the action set at time step $n$ consists of two separate actions:

\textbf{Action 1 (for selecting a split option)}: Given a selected split option at the previous time step, an action is taken at time step $n$ out of the following three options (based on the latency requirement and traffic flow of the selected split option):
\begin{itemize}
    \item Moving RAN split point towards the first-upper split point
    \item Moving RAN split point towards the first-lower split point
    \item No action is taken,
\end{itemize}
\textcolor{black}{in which the split point is where the network functions is splitted to DU functions and the CU functions (Fig. 2).} The reason behind such a definition is twofold. First, this definition can simply show the impacts of the split point and split option on the traffic flow and latency requirement, i. e., based on the Fig. 2, moving split point towards upper-layer points at the protocol stack can usually result in lower traffic demand but higher latency, and vice versa. Second, defining this action with only three options leads to a small action space size.

\textbf{Action 2 (for selecting a RAN option)}: Assuming a selected RAN for the UE access in the previous time step, i. e., LEO satellite-based O-RAN or HAPS-based O-RAN, at time step $n$, Action 2 would be selecting one of the following two options:
\begin{itemize}
    \item Keeping access to the current O-RAN
    \item Selecting another O-RAN.
\end{itemize}
Assuming three options for Action 1 and two options for Action 2, the total action space is
6 which is a reasonable size for the DQN method. 

\subsection{Reward}

To fully characterize network performance and provide a correct trajectory for the DQN method, the reward function in the time step $n$ is defined as follows
\begin{equation}
    Reward(n)=\sum_{j=1}^{4}\alpha_{j}R_{j}(n),
\end{equation}
where
\begin{equation}\label{rew1}
    R_{1}(n)=\begin{cases}
        +1,& L_{Total}^{o(n)}\leq L^{o(n)}, \\
        -1,& L_{Total}^{o(n)}> L^{o(n)},
    \end{cases}
\end{equation}
\begin{equation}
    R_{2}(n)=\begin{cases}
        +1,& TRA^{o(n)}\leq C_{i(n),GAT}, \\
        -1,& TRA^{o(n)}>C_{i(n),GAT},
    \end{cases}
\end{equation}
\begin{equation}
    R_{3}(n)=\begin{cases}
        +1,& {COMP_{i(n)}^{o(n)}}\leq {COMP_{max,i(n)}}, \\
        -1,&{COMP_{i(n)}^{o(n)}}> {COMP_{max,i(n)}},
    \end{cases}
\end{equation}
\begin{equation}
    R_{4}(n)=\begin{cases}
        +1,& {COMP_{GAT}^{o(n)}}\leq {COMP_{max,GAT}}, \\
        -1,&{COMP_{GAT}^{o(n)}}>{COMP_{max,GAT}},
    \end{cases}
\end{equation}
while $\alpha_{j}, \ j\in\{1,2,3,4\}$ are penalty positive-value coefficients regarding $R_{1}(n)$ to $R_{4}(n)$ parameters that are set during the training process with appropriate weights. 
It is noteworthy that selecting options for Action 1 and Action 2 results in setting a specific value for $y_{i}^{o}$ in the time step the action is taken. The obtained value for this variable will be assessed in each time step based on the reward defined above.

\section{Numerical Analysis}
\begin{figure}[t]
    \centering
    \includegraphics[width=0.80\columnwidth]{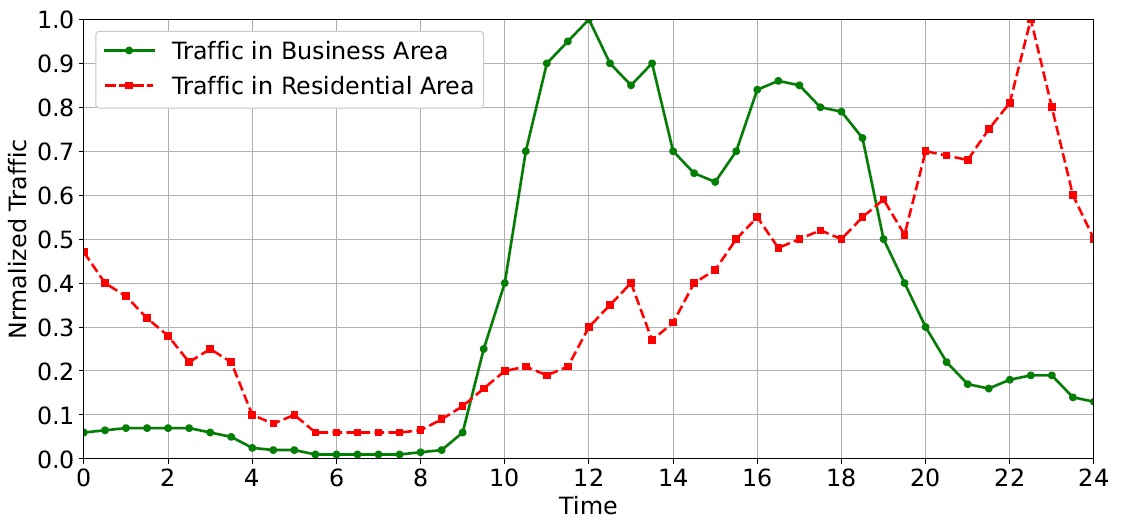}
    \caption{Daily traffic pattern in a residential area and a business area  for weekdays based on measurements in \cite{marsan2013towards}.}
    \label{Traffic Model}
\end{figure}

\begin{figure}
    \centering
    \includegraphics[width=0.5\columnwidth]{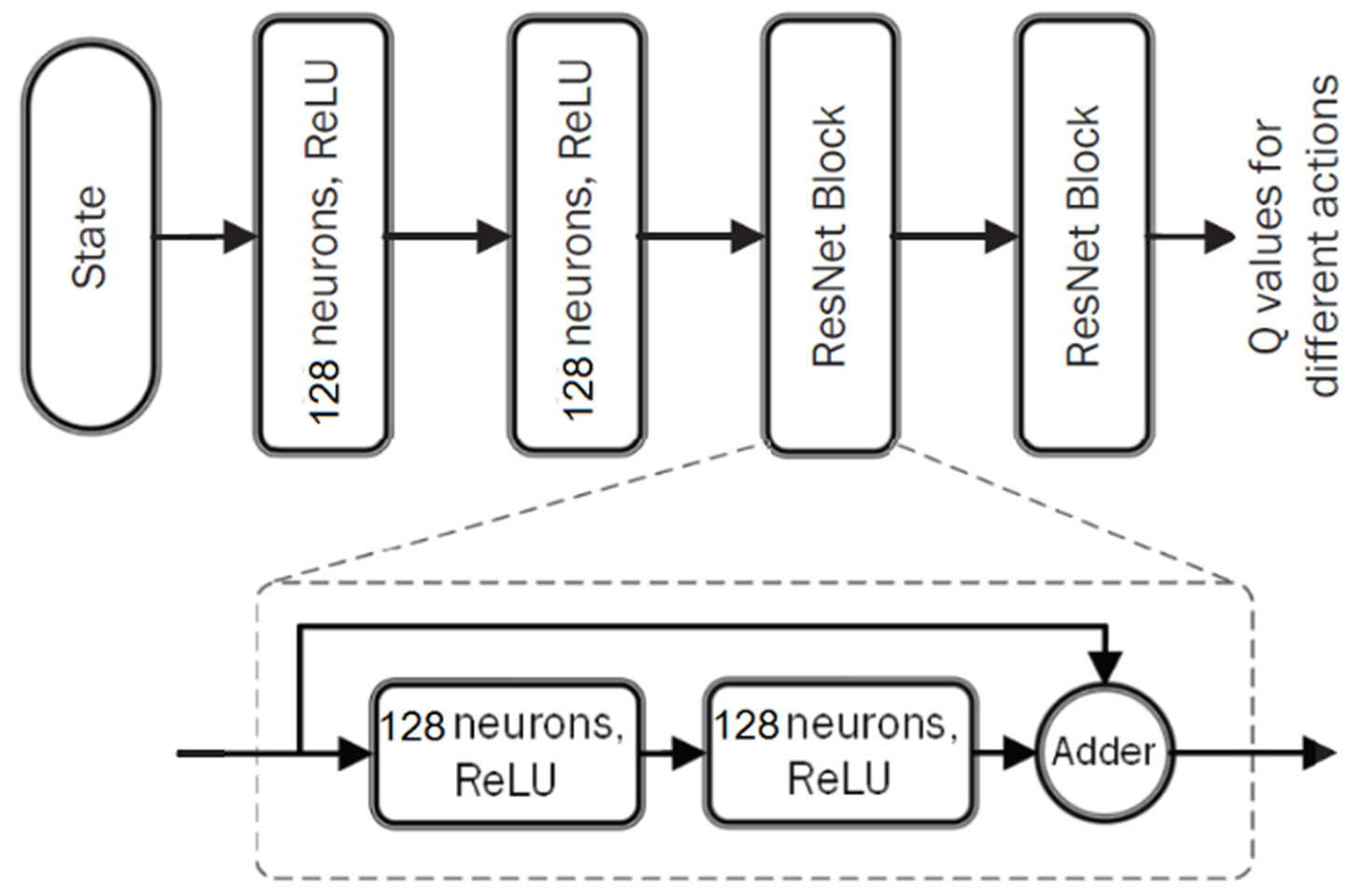}
    \caption{ResNet-aided QNN structure}
    \label{ResNet QNN}
\end{figure}
In this section, the performance of the proposed method is evaluated according to different traffic loads and latency requirements in downlink transmission of the feeder link. In order to apply a dynamic RU traffic flow $\lambda_{RU}$ to the system model, the daily traffic pattern introduced in \cite{marsan2013towards} is utilized for both business area and residential area  within weekdays with the peak traffic of 200 Mbps and average traffic  of 100 Mbps, and the corresponding RU traffic pattern is shown in Fig. \ref{Traffic Model}. The normalized power consumption is considered as the performance benchmark. For the sake of a comparative analysis, the Seq-to-Seq functional split method  in \cite{amiri2023energy},  which has been designed for load balancing and traffic routing in the 5G terrestrial networks, is  extended and applied to our NTN-based O-RAN scenario, and this extended method is compared with our proposed method. The parameters of the NTN nodes and the gateway  are shown in Table III. The values of $EPO$s, computational capacities and idle powers in the table come from Nvidia Jetson TX2, Nvidia Jetson AGX Xavier and Nvidia L4 used for processing units of the HAPS, LEO satellite and the gateway, respectively. 
In this work, we assume the energy consumption of the NTN nodes does not exceed the battery discharge threshold energy during the time the NTN platform is serving the cell. However, as a future work, we will consider cases in which there are some limitations in this regard. 
\begin{table}[t]
\vspace{15pt}
\caption{Simulation Parameters \cite{3gpp2023solution},\cite{lou2023haps} and \cite{matoussi20205g}}
    \label{tab:my_label}
    \centering
    \begin{tabular}{c|c}
        Parameter & Value \\
        \hline
        $COMP_{PHY}$ &  1280 GOPS  \\
        $COMP_{RLC}$ &  50 GOPS  \\
        $COMP_{MAC}$ &  50 GOPS \\
        $COMP_{PDCP}$ &  100 GOPS  \\
        $COMP_{max,i}$ & $SAT$: 32 TOPS, $HAP$: 1.33 TOPS   \\
        $COMP_{max,GAT}$ & 485 TOPS  \\
        $d_{i,GAT}$  &  $SAT$: 600 km, $HAP$: 20 km     \\
        $EPO_{i}$  &   $SAT$: 0.625 J/TO, $HAP$: 5,64 J/TO  \\

        $EPO_{GAT}$  & 0.0742 J/TO
        \\
        $P_{I,i}$  & $SAT$: 10 W , $HAP$: 7.5 W      \\
        $P_{I,GAT}$  &    36 W  \\
        $C_{i,GAT}$ &  $SAT$: 100 Mbps, $HAP$: 10 Gbps \\ 
        $p_{i,GAT}$ &  $SAT$: 35 W, $HAP$: 4 W \\
    \end{tabular}
\end{table}
 The DQN architecture employs a deep residual network (ResNet) \cite{he2016deep} with six hidden layers, as outlined in Fig. \ref{ResNet QNN}. Each of these hidden layers consists of 128 neurons. To simplify, we denote this ResNet-based Q-neural network (QNN) as simply ResNet. Within this structure, each component is termed a ResNet block. The activation functions for the neurons are ReLU functions. The initial two hidden layers of the ResNet are fully connected, succeeded by two ResNet blocks. Each ResNet block comprises two consecutive hidden layers alongside a shortcut linking the input to the output of the block. When updating the coefficients in the QNN, a mini-batch comprising 100 experimental samples is randomly chosen from the experience replay reservoir containing the 200 previous experiences to compute the loss function. Notably, the experience replay reservoir follows a first-in-first-out (FIFO) approach, with older experiences replaced as the reservoir reaches capacity. The RMSprop algorithm \cite{goodfellow2016deep} can be employed to update the weights of the QNN using the mini-batch gradient descent method. Additionally, an exponential Decay $\epsilon$-greedy algorithm is implemented to prevent the DRL structure from becoming trapped in a sub-optimal decision policy before accumulating sufficient experiences. While $\epsilon$ is initially set to 0.5, it is decreased gradually
with every time step by the rate of 0.995, until it reaches a threshold of 0.0005. 
\begin{figure}[t]
    \centering
    \includegraphics[width=0.94\columnwidth]{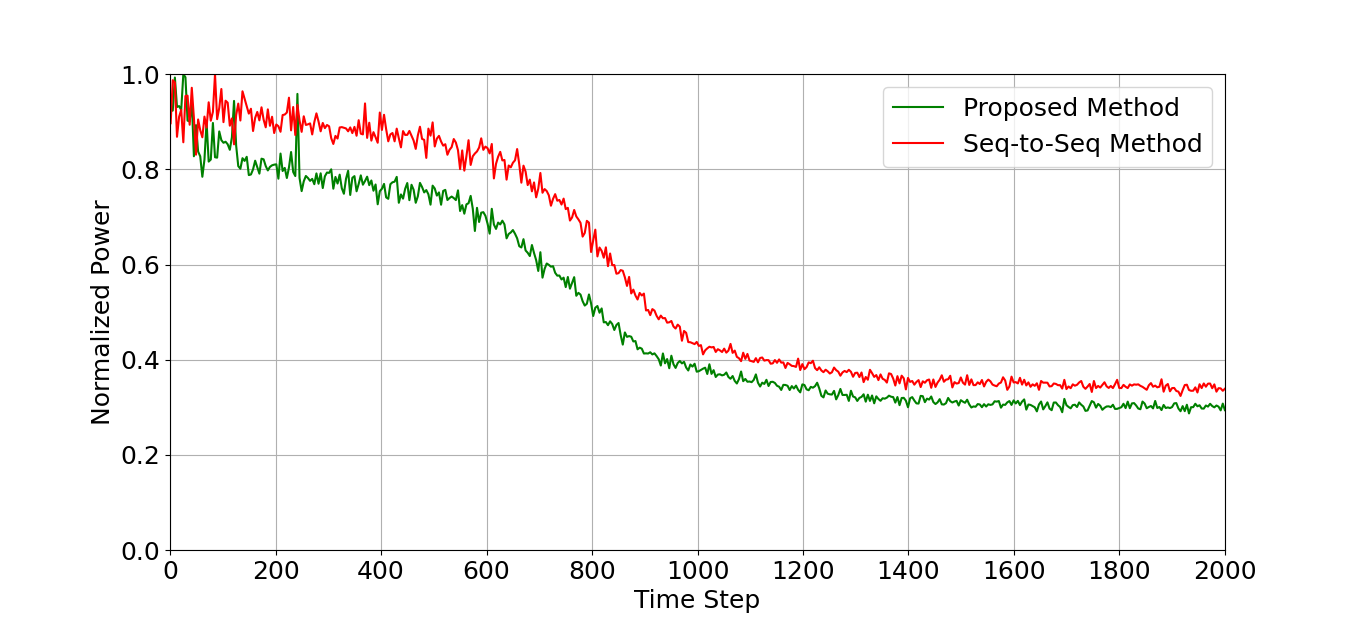}
    \caption{Normalized power vs time step for a business area}
    \label{NORPoWBus}
\end{figure}

\begin{figure}[t]
    \centering
    \includegraphics[width=0.94\columnwidth]{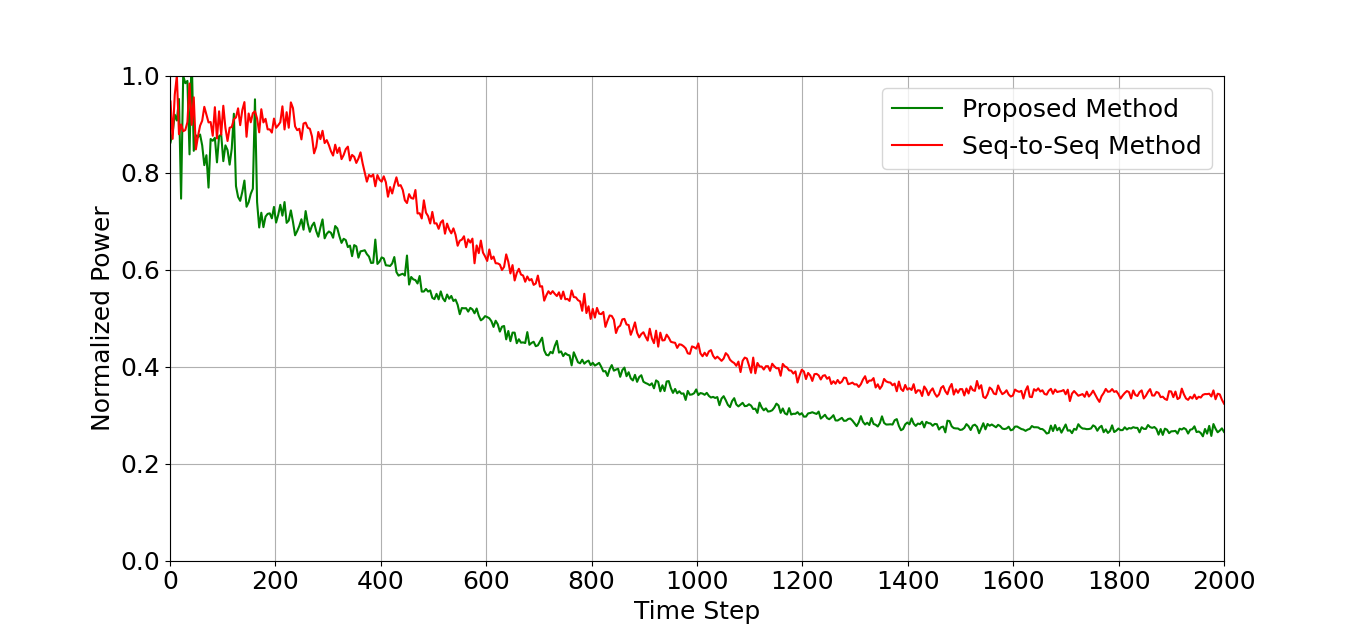}
    \caption{Normalized power vs time step for a residential area}
    \label{NorPowRes}
\end{figure}
As it is shown in Fig. \ref{NORPoWBus}, for a business area, after selecting a random action set, i. e., selecting a split option and a RAN network, that meets the optimization constraints, the DQN agent is able to adopt a correct trajectory, and revise its actions after some time steps accordingly based on the feed-back obtained from the network environment through the rewards it gains. Also it is evident while the proposed method outperforms the Seq-to-Seq method in terms of normalized power consumption by 20 percents, it provides a faster convergence. The similar results can be found for residential area as shown in Fig. \ref{NorPowRes} which implies the efficiency and the effectiveness of the proposed method  for NTN-based O-RAN architectures. Fig. 7 illustrates the  adaptability of the proposed platform to varying traffic patterns, showing selected split options for different times of a day. As depicted, during peak traffic hours, e. g., around 8 am for the  business area and around 10 pm for the residential area, the split option 3 is selected due to its ability to support higher traffic load. Outside of these peak periods, different split options are selected according to latency requirements, traffic load and computational load coming from the network at various times of the day.
\begin{figure}[t]
    \centering
    \includegraphics[width=0.6\columnwidth]{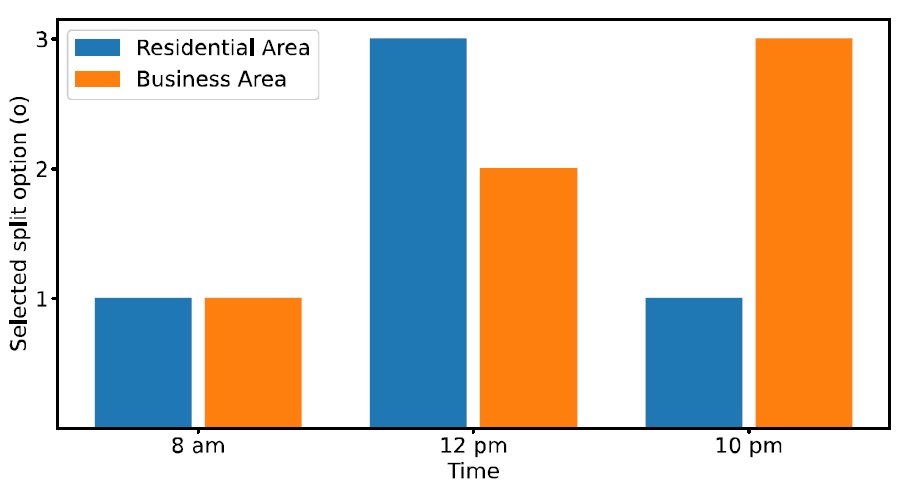}
    \caption{The selected split option for different times of a day}
    \label{fig:enter-label}
\end{figure}
\section{Conclusion}
This study introduced a DQN-based reinforcement learning framework to optimize the functional split in O-RAN integrated with NTN networks. Our approach dynamically selects optimal RAN configurations, showing significant improvements in energy efficiency across various platforms like LEO satellites and HAPSs. Moreover, the proposed DQN-based method demonstrates the capability to adjust to diverse traffic flows derived from daily traffic patterns. It efficiently selects the optimal RAN functional split option and the NTN platform from the available options.

\bibliographystyle{IEEEtran}
\bibliography{main}

\end{document}